\documentclass[aip,amsmath,amssymb,reprint,floatfix,longbibliography,jcp]{revtex4-2}
\usepackage{graphicx}

\newcommand{\odiv}{\operatorname{div}}
\newcommand{\br}{\ensuremath{\mathbf{r}}}

\newcommand{\rT}{\ensuremath{\mathrm{T}}}
\newcommand{\bx}{\ensuremath{\mathbf{x}}}
\newcommand{\by}{\ensuremath{\mathbf{y}}}
\newcommand{\bz}{\ensuremath{\mathbf{z}}}
\newcommand{\bbee}{\ensuremath{\mathbf{b}}}

\newcommand{\htwo}{\ensuremath{\mathrm{H}_2}}
\newcommand{\htwospace}{\ensuremath{\mathrm{H}_2}\ }

\usepackage[table,xcdraw,dvipsnames]{xcolor}
\usepackage{makecell}

\begin{document}

\title{Learning Memory-Dependent Neural Network Correlation Potentials for Accurate Electron Dynamics}

\author{Harish S. Bhat}
\email{hbhat@ucmerced.edu}
\affiliation{Department of Applied Mathematics, University of California Merced, Merced, California 95343, USA}

\author{Christine M. Isborn}
    
    \affiliation{Department of Chemistry and Biochemistry, University of California Merced, Merced, California 95343, USA}

\date{31 August 2026} 

\begin{abstract}\label{abstract}
Time-dependent density functional theory, though exact in theory, is in practice applied in an adiabatic approximation using exchange-correlation functionals with only local temporal dependence.  Simultaneously, the exact correlation potential formally depends on, among other quantities, the time-history of electron densities.  Here we develop a framework to learn neural network models of correlation functionals that feature explicit memory-dependence.  Our framework includes two different approaches, both linked through their use of adjoint-based optimization.  One approach decouples the learning of the functional from inversion to find ground truth values of the correlation potential.  The other approach learns the functional directly without requiring inversion.  We apply these methods to modeling the electron dynamics of two-electron systems in two spatial dimensions.  In both cases, our methods yield correlation functionals with low test set propagation error, outperforming standard local density and generalized gradient approximation functionals by one to two orders of magnitude.  Overall, our framework points at one strategy to move beyond the adiabatic approximation and develop memory-dependent correlation functionals that yield accurate propagation for excited state and/or non-equilibrium dynamics.
\end{abstract}

\maketitle

\section{Introduction}
\label{sect:intro}
Electronic evolution is essential for many processes, from charge transfer to energy capture to nonequilibrium chemical phenomena. Time-dependent density functional theory (TDDFT) is considered the workhorse for modeling molecular electronic excited states and excited state properties within electronic structure theory due to its balance of accuracy and computational efficiency.\cite{FA02,Jacquemin2009,Jacquemin2011,Laurent2013,Adamo2013,OBFS15,SBMLJ21,Liang2022,Mester2022}  Although formally exact, and often applied to simulate electron dynamics, \cite{Provorse2016b, Li2020,Herbert2023, Xu2024} TDDFT is almost always applied within an adiabatic approximation, in which the exchange-correlation potential only takes as input the properties of the instantaneous electron density.\cite{Runge1984, GK90, ullrich2011time, WYB12, GM12chap, MBAG02, M16, Ullrich2025} This adiabatic approximation ignores the density at all previous points in time, and the potential therefore has no memory of the past state or density of the system,\cite{Maitrachap} leading to inaccurate electronic evolution, including incorrect descriptions of Rabi oscillations,\cite{Habenicht2014} missing states with double excitation character,\cite{Maitra2004, M22} incorrect pole structure in the quadratic response,\cite{Parker2016, Dar2023} spuriously shifting peaks and resonances,\cite{Provorse2015, FLSM15, LFM16} and incomplete charge transfer due to missing dynamical, sharp features in the potential.\cite{LEM13, LFSEM14, M17,M22} Although the failures of the adiabatic approximation are likely more extreme for smaller systems,\cite{Ranka2023} and may be improved by a response reformulation based on linear response theory,\cite{DBM24} accurate TDDFT electron dynamics far from the ground state requires new approaches. Going beyond an adiabatic approximation to the potential requires learning the memory-dependence that should be built into the potential.

TDDFT can in principle produce exact electron density dynamics by evolving non-interacting electrons within a one-body potential. These electrons generally reside within  non-interacting Kohn-Sham (KS) orbitals $\{ \varphi_j(\bx, t)\}$ that satisfy the time-dependent Kohn-Sham (TDKS) equation, stated here in atomic units:
\begin{equation}
\label{eqn:tdks}
i \partial_t \varphi_j = \left( -\frac{1}{2} \nabla^2 + v^S \right) \varphi_j,
\end{equation}
with the one-body density $\rho$ obtained by summing over occupied KS orbitals $\rho = \sum_j| \varphi_j |^2$.  Here the KS potential satisfies
\begin{equation}
\label{eqn:vS}
v^S = v^\text{ext} + v^H + v^X + v^C,
\end{equation}
where the potentials on the right-hand side are, respectively, external, Hartree, exchange, and correlation potentials. The correlation potential is unknown and must be approximated. It is known that the correlation potential $v^C$ formally depends on the initial KS state $\varphi_0$, the initial interacting wave function $\Psi_0$, and the time-history of electron densities $\rho_t$, and thus has memory of the system at previous points in time. Attempts have been made to add this memory-dependence to the potential, with focus on the frequency-dependence of the exchange-correlation kernel within the linear response regime\cite{Gross1985, Maitra2004, Panholzer2018, Kaplan2022} and through adding dependence on the current-density.\cite{Vignale1996, Dobson1997} 

Alternatively, if near-exact electron densities are available, the time-history and spatial nonlocality of the density, can be incorporated into the potential by either directly learning a functional to accurately propagate the density or by generating accurate training data to learn such a functional. For a two-electron system in one spatial dimension, we have applied the direct learning approach previously by applying adjoints to develop efficient methods to compute gradients and numerically solving the time-dependent Schr\"odinger equation (TDSE) to obtain electron densities suitable for training machine learning models of the correlation potential.\cite{Bhat2022} Note that with this direct learning method, there is no ground truth correlation potential involved in the learning, only ground truth densities. Alternatively, for a two-electron system, the TDKS equations can be inverted using the quantum hydrodynamics\footnote{We explain in Section \ref{sect:ilh2} why we use this term.} approach explained in Appendix E of \citet{ullrich2011time}.  This inversion yields spacetime values of the correlation potential, which can be used to machine learn a correlation functional.  This invert-then-learn machine learning approach that can build memory into the correlation potential has been pursued for a spatially one-dimensional electron-hydrogen scattering problem by Suzuki et al.\cite{PhysRevA.101.050501} Inversion beyond one spatial dimension has been carried out by the Maitra group, where dynamical step and peak features in the potential were found for the 3D helium atom.\cite{DLFM21}  However, in this latter work, the computed $v^C$ was not used to propagate the TDKS equations, and therefore the accuracy of the resulting electronic evolution is unknown. For systems with more than two electrons that require multiple Kohn-Sham orbitals, quantum hydrodynamics no longer furnishes a solution of the inversion problem. Thus, it is worth pursuing a more general constrained optimization approach to inversion and learning, that, in principle, could be generalized to many-electron systems. 

In this paper, we use a time series of exact 1-electron densities $\widetilde{\rho}$ from a two-electron system to learn memory-dependent neural network models of the unknown correlation potential $v^C$.  We explore two different and complementary approaches. For the first, we invert the TDKS equations to solve for the spacetime values of the correlation potential such that when we solve the TDKS system (\ref{eqn:tdks}) using this $v^C$,  the resulting solution yields a 1-electron density that minimizes the mismatch between $\rho$ and the ground truth value, $\widetilde{\rho}$. For the second, we employ direct learning to solve for a \emph{functional} $v^C[\rho_t; \theta]$, parameterized by $\theta$.  For this latter direct learning problem, we have the ability to find one functional that minimizes the mismatch between \emph{multiple} $\rho$ trajectories and their ground truth counterparts $\widetilde{\rho}$. We show how to couple constrained optimization and numerical solution of (\ref{eqn:tdks}) with neural network models of the correlation potential.

For conventional ground-state DFT (density functional theory), significant efforts have been made to learn exchange-correlation functionals.\cite{Tozer1996,nagaicompleting,Schmidt2019,dick2020machine,Cuierrier2021,Kasim2021,Kirkpatrick2021,Burke2021,margraf2021pure,Bystrom2022,Kalita2022,liu2023supervised,Bystrom2024,akashi2025machineslearndensityfunctionals,polak2025real,gao2025learning}  Much of this work follows an invert-then-learn approach: (i) accurate reference densities $\rho$ are computed (typically via wave function methods), (ii) these densities are fed into a Kohn-Sham inversion procedure that yields ground truth values of $v^{XC}$, and (iii) supervised learning is applied to develop a functional $v^{XC}[\rho]$.  A notable recent extension of this procedure augments $v^{XC}$ with reference exchange-correlation energies $E^{XC}$ (computed using results of Kohn-Sham inversion and wave function data) to learn an $E^{XC}[\rho]$ functional constrained such that both $E^{XC}[\rho]$ and $v^{XC}[\rho] = \delta E^{XC} / \delta \rho$ match their respective reference values.\cite{Kanungo2025}  We also find recent work that directly learns a functional without employing Kohn-Sham inversion\cite{luise2026accuratescalableexchangecorrelationdeep}; here the data consists of reference densities $\rho$ together with numerous physical observables across a large number of systems.  Because ground-state DFT is a time-independent theory, the issue of memory-dependence of $v^{XC}$ never arises.

Specifically within the context of TDDFT, to our knowledge, this is the first work to treat the problem of machine learning a correlation functional in more than one spatial dimension.  The invert-then-learn approach we pursue here enables rapid training of models with long memory (in this case, up to $M \Delta t =1.28$ a.u.); these models can then be fine-tuned using adjoint-based methods.  The fine-tuned neural network yields a trained correlation functional that reproduces the large spatial gradients and non-local features present in the correlation potential  obtained through highly accurate PDE-constrained optimization/inversion.  The direct learning approach, in which the neural network never sees ground truth or reference values of the correlation potential, yields $L_2$ propagation error less than $3 \times 10^{-3}$ on the test set despite being trained only on two trajectories.   Overall, our trained neural network functionals yield test set propagation errors that are one to two orders of magnitude less than those obtained via standard local density and generalized gradient density functionals.  
The methodological framework we establish is general and can be extended to larger training sets and three-dimensional systems with no substantive changes.

\section{Methods}
Let us consider TDDFT for two-electron systems in two spatial dimensions.  We will model these systems using a single, doubly occupied KS orbital $\varphi(\bx, t)$ with $\rho = 2 | \varphi |^2$.  As we have only one doubly-occupied KS orbital, the exact exchange potential is $-1/2$ the Hartree potential.  Using this, we can write
\begin{equation}
\label{eqn:vS2}
v^S[\rho_t, \varphi_0, \Psi_0] = v^\text{ext} + \frac{1}{2} v^H[\rho] + v^C[\rho_t, \varphi_0, \Psi_0].
\end{equation}
Let $V_{\alpha}(r) = 1/\sqrt{ r^2 + \alpha^2}$ denote the soft-Coulomb potential with parameter $\alpha \geq 0$.  When $\alpha = 0$, we obtain $V_0(r) = 1/r$, the standard Coulomb potential.  In terms of this, we can write the Hartree potential as
\begin{equation}
\label{eqn:Hartree}
v^H[\rho](\bx, t) = \int_{\Omega} \rho(\by, t) V_{\alpha}(| \bx - \by |) \, d \by,
\end{equation}
where $| \bx - \by |$ is the Euclidean distance between $\bx$ and $\by$. For each of the model problems considered below, the external potential $v^\text{ext}$ in (\ref{eqn:vS2}) is known. The only unspecified term on the right-hand side of (\ref{eqn:vS2}) is the correlation potential $v^C$.

To proceed, it is natural to discretize the TDKS problem in space and time.  We choose a split-step discretization of (\ref{eqn:tdks}).   Each time step consists of a half-step of the kinetic propagator, a full step of the potential propagator, and a second half-step of the kinetic propagator. Let $\hat{\rT} = (-1/2) \nabla^2$ denote the kinetic operator and $\rT$ a particular spatial discretization thereof.  Then we can formulate our discretization of (\ref{eqn:tdks}):
\begin{multline}
\label{eqn:forwardprop}
\varphi^{k+1} =  \exp(-i \Delta t \rT/2) \\ 
\cdot \exp \biggl[-i \Delta t V^S_k[\rho; \theta] \biggr] \exp(-i \Delta t \rT/2) \varphi^k.
\end{multline}
Here each $\varphi^k$ is a complex $N^2 \times 1$ vector. When we take the continuous space-time KS potential $v^S$ and evaluate it on our spatial grid at time $k \Delta t$,
we obtain an $N \times N$ matrix, which we denote $V^S_k$.  We will use this capitalization consistently, so that $V^\text{ext}_k$, $V^H_k$, and $V^C_k$ denote matrices formed by evaluating the corresponding $v^\text{ext}$, $v^H$, and $v^C$ potentials on spatial grids at time $k \Delta t$. Let us abbreviate the kinetic and potential propagators:
\begin{align}
\label{eqn:Kdef}
K &= \exp(-i \Delta t \rT/2) \\
\label{eqn:Vpropdef}
V^\text{prop}_k [\rho^k; \theta] &= \exp(-i \Delta t V_k^S[\rho^k; \theta] ).
\end{align}
If we think of $\varphi^k$ as a complex vector of size $N^2 \times 1$, then these propagators are matrices of size $N^2 \times N^2$.  In this representation, $V^\text{prop}_k$ is a purely diagonal matrix.  Details of kinetic propagator computation will be given in Section \ref{sect:compdetails}.

The goal of this paper is to use time series  of 1-electron densities to learn memory-dependent neural network models of the correlation potential $v^C$.  As we detail in Sections \ref{sect:htwo} and \ref{sect:scattering}, for our two model systems, we solve the TDSE to generate reference data $\{ \widetilde{\rho}^k \}_{k=0}^{N_s}$ on a spatiotemporal grid. Here each $\widetilde{\rho}^k$ is a real, $N \times N$ matrix, corresponding to the values of the (nonnegative) density on an equispaced grid in two-dimensional space.  The tilde on the $\widetilde{\rho}$ signifies that this is \emph{ground truth} or \emph{reference} data computed from solutions of the TDSE (e.g., not computed via TDKS/TDDFT). 

We devise two approaches to achieve our goal. For clarity of exposition, we first formulate both in continuous time and space, assuming access to ground truth densities $\widetilde{\rho}(\bx, t)$ defined on $0 \leq t \leq T$ and $\bx \in \Omega$.

The \emph{invert-then-learn} approach proceeds in two stages.  In the first stage (inversion), we solve for the spacetime \emph{values} $v^C(\bx, t)$ of the correlation potential such that when we solve the TDKS system (\ref{eqn:tdks}) using this $v^C$,  the resulting solution $\varphi$ yields a 1-electron density $\rho = 2 | \varphi |^2$ that minimizes the mismatch between $\rho$ and $\widetilde{\rho}$, measured by the squared $L_2$ norm in space, integrated over time:
\[
\frac{1}{2} \int_0^{T} \int_{\Omega} ( \rho(\bx, t) - \widetilde{\rho}(\bx, t) )^2 \, d \bx \, dt.
\]
In the inversion problem, each time we change the ground truth $\widetilde{\rho}$, the optimal correlation potential $\widetilde{v}^C$ (evaluated pointwise in spacetime) also changes. Equipped with enough pairs of the form $(\widetilde{\rho}, \widetilde{v}^C)$, we can machine learn a functional that maps time histories of densities to correlation potentials.

The \emph{direct learning} problem is identical to the inversion problem except that we solve for a \emph{functional} $v^C[\rho_t,\varphi_0,\varphi_0; \theta]$, parameterized by $\theta$.  The only other difference is that in the direct learning problem, we have the ability to find one functional that minimizes the mismatch between \emph{multiple} $\rho$ trajectories and their ground truth counterparts $\widetilde{\rho}$.  

In both inversion and learning, we can simplify notation by allowing $V^C_k$ and thus $V^S_k$ to depend on a set of parameters $\theta$.  In the inversion case, this allows one to represent $V^C$ as a contraction of $\theta$ against discretized basis functions.  In this setting, our choice is to let $\theta$ be an $N_s \times N \times N$ tensor, in which case $V^C_k = \theta_{k, :, :}$---the elements of $\theta$ are the values of the potential, or equivalently, we have represented $V^C$ in the discrete position basis.   With this common framework, we can formulate and solve both inversion and learning as optimization problems constrained by the discretization of the TDKS equations.

The adjoint method underpins both the invert-then-learn and direct learning strategies.  Essentially, our loss function in both cases measures the error incurred by solving (\ref{eqn:tdks}) with a fixed choice of parameters $\theta$ that determine either the values or the functional form of the correlation potential.  Our formulation shares much in common with quantum optimal control, where adjoint methods have been used extensively.  \cite{Rabitz1988,borzi2017formulation,BhatTDHFOC2025,BhatQOCHessian}  We pursue the adjoint method here because (i) it enables calculation of exact gradients at a computational cost that scales (with problem size) at the same rate as TDKS propagation itself, (ii) almost all terms needed to evaluate the adjoint systems (\ref{eqn:genadj}) and (\ref{eqn:genadjlearn}) are computed during TDKS propagation, with the remaining ones easily evaluated via automatic differentiation, and (iii) direct automatic differentiation of the long-term propagation error fails to exploit the particular structure of TDKS propagation, leading to excessive computational effort.

\subsection{Inversion: Adjoint Method}
\label{sect:invadj}
The first step of the invert-then-learn method is to carry out partial differential equation (PDE)-constrained optimization to solve for reference values of the correlation potential.  A more precise formulation of the inversion problem is to find $\theta$ to minimize the loss
\begin{equation}
\label{eqn:cost}
C(\theta) = \frac{1}{2} \sum_{j=1}^{N_s} \| \rho^j - \widetilde{\rho}^j \|^2
\end{equation}
subject to the discretized equations of motion 
\begin{equation}
\label{eqn:disceom}
\varphi^{k+1} = K V^\text{prop}_k(\rho^k; \theta) K \varphi^k.
\end{equation}
We assume that we work with a fixed TDKS initial condition $\varphi^0$ such that $\rho^0 = 2 \varphi^0 (\varphi^0)^\ast = \widetilde{\rho}^0$. We form a Lagrangian that incorporates the loss and dynamical constraints:
\begin{subequations}
\label{eqn:Lagrangian}
\begin{align}
&L(\varphi,\Lambda,\theta) = L_1(\varphi,\Lambda,\theta) + L_2(\varphi,\Lambda,\theta) \\
&L_1(\varphi,\Lambda,\theta) = \frac{1}{2}\sum_{j=1}^{N_{\text{s}}} (\rho^j - \widetilde{\rho}^j)^\dagger (\rho^j - \widetilde{\rho}^j)\\
\begin{split}
&L_2(\varphi,\Lambda,\theta) \\
&= -\Re \sum_{j=0}^{N_{\text{s}}-1} (\lambda^{j+1})^\dagger (\varphi^{j+1}  - K V_j^\text{prop}[\rho^j; \theta] K \varphi^j ).
\end{split}
\end{align}
\end{subequations}
Here $\varphi = \{ \varphi^j \}_{j=1}^{N_s}$ is the collection of all TDKS states, while $\Lambda = \{ \lambda^j \}_{j=1}^{N_s}$ is the collection of all adjoint states.  Lagrange multiplier theory tells us that a minimizer of the constrained problem must be a critical point of the Lagrangian.

In our search for a critical point, let us begin with the gradient of $L$ with respect to the adjoints $\Lambda$. Setting this to zero, we recover our equations of motion (\ref{eqn:disceom}). We move on to take variations of $L$ with respect to both $\varphi$ and $\varphi^\ast$.  
\begin{figure*}[t]
\includegraphics[width=6in,clip,trim=50 150 50 10]{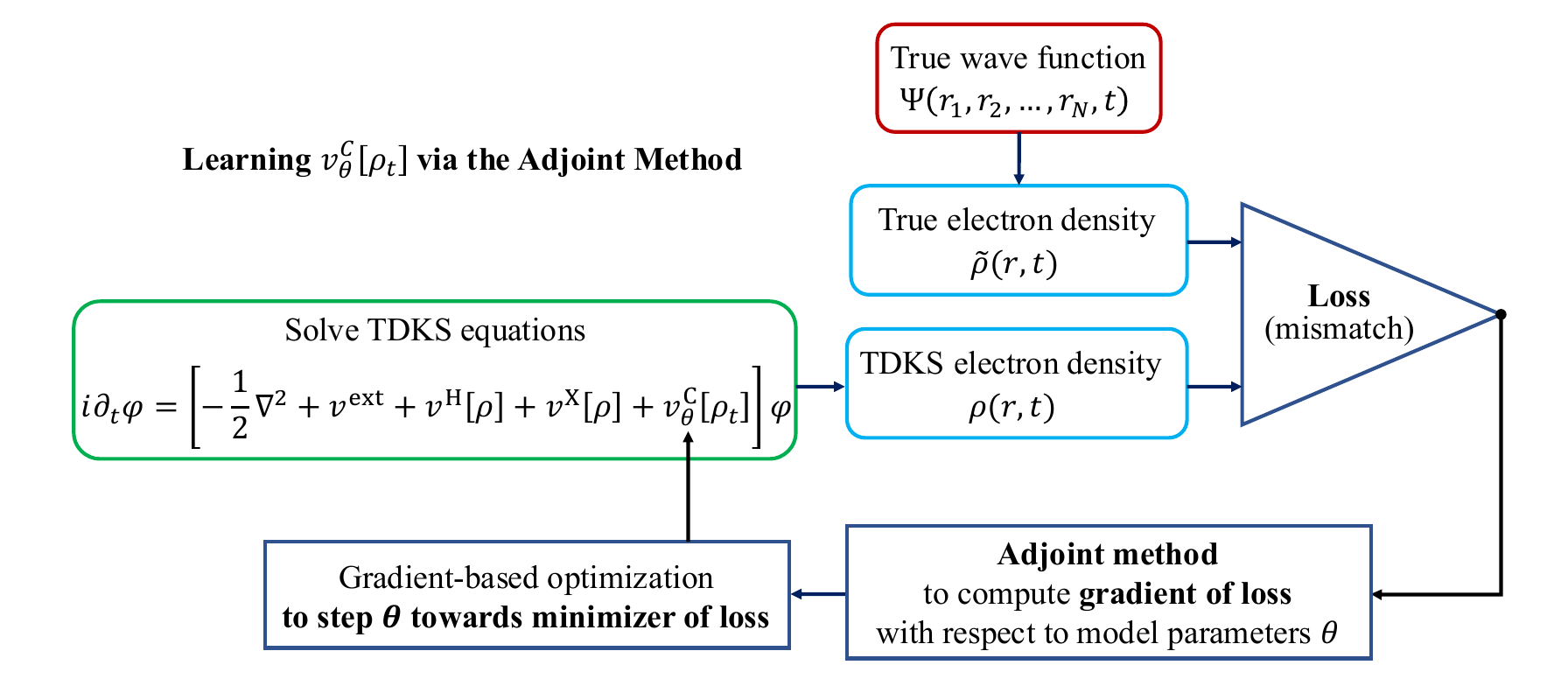}
\caption{This diagram gives a high-level overview of how we employ the adjoint method to solve either the inversion or direct learning problems.  Both involve parameterizing the unknown correlation potential $v^C$ with parameters $\theta$.  On each iteration, we \emph{propagate} the TDKS equations with $v^C$ evaluated using our current iterate of $\theta$, enabling us to compute the loss (the mismatch between true and TDKS densities).  To minimize the loss, we must be able to compute its gradient.  The adjoint method is designed to compute this gradient in an accurate, scalable way.}
\label{fig:adjdiag}
\end{figure*}
To compute this, we will use variational notation. We have two options here: (i) take variations with respect to the real and imaginary parts of $\varphi$, separately, or (ii) take variations with respect to both $\varphi$ and $\varphi^\ast$.  Both (i) and (ii) are mathematically equivalent.  We choose option (ii) as it is notationally simpler.  We begin by noting that $\rho = 2 | \varphi |^2$ implies
\[
\delta \rho^j = 2 \varphi^j (\delta \varphi^j)^\ast + 2 (\varphi^j)^\ast \delta \varphi^j = 4 \Re [ (\varphi^j)^\ast \delta \varphi^j ].
\]
With this,
\[
\delta L_1 = \sum_{j=1}^{N_s} (\rho^j - \widetilde{\rho}^j )^\dagger \delta \rho^j = \Re \sum_{j=1}^{N_s} 4 (\rho^j - \widetilde{\rho}^j )^\dagger (\varphi^j)^\ast \delta \varphi^j.
\]
Having computed the variation of $L_1$, we turn to the variation of $L_2$, but first we note that
\[
\delta \varphi^{k+1} = K \delta V_k(\rho^k; \theta) K \varphi^k + K V_k(\rho^k; \theta) K \delta \varphi^k.
\]
As discussed above, we can treat $V_k^\text{prop}$ as a diagonal matrix, in which case its $(a,a)$-th element is
\[
V^\text{prop}_{k; aa}(\rho^k; \theta) = e^{ -i \Delta t \bigl(V^{\text{ext}}_a + \frac{1}{2} V^H[\rho^k]_a + V^C_a(t_k; \theta) \bigr) }
\]
where $a$ goes from $1$ to $N^2$.  The Hartree potential depends functionally on the electron density $\rho$.  We can equivalently think of $V^H[\rho^j]$ as an $N \times N$ matrix, which gives the value of the potential at each two-dimensional spatial grid point---or as a flattened vector of length $N^2$.  In the expression $V^H[\rho^j]_a$, we are implicitly choosing the flattened representation.  We use ${\partial V^H_a}/{\partial \rho^j}$ to denote the $a$-th component of the derivative of $V^H[\rho]$ with respect to $\rho = \rho^j$.  Then
\begin{multline}
\label{eqn:deltaV}
\delta V^\text{prop}_{j; aa}(\rho^j; \theta) = V^\text{prop}_{j;aa}(\rho^j; \theta) (-i \Delta t) \left( \frac{1}{2} \frac{\partial V^H_a}{\partial \rho^j} \delta \rho^j  \right) \\
= V^\text{prop}_{j;aa}(\rho^j; \theta) (-i \Delta t) \left[ \frac{\partial V^H_a}{\partial \rho^j} ( (\varphi^j)^\ast \delta \varphi^j + (\delta \varphi^j)^\ast \varphi^j )  \right].
\end{multline}
The remaining details of the derivation of the variation $\delta L_2$ can be found in supplementary material.  Here we note that after $\delta L_2$ has been computed, one can combine results to obtain $\Delta L$.  Setting $\delta L = 0$ for all variations of $\varphi$ and $\varphi^\ast$, we obtain the following backward-in-time system. At $j = N_{\text{s}}$, we have
\begin{equation}
\label{eqn:fincond}
(\lambda^{N_{\text{s}}})^\dagger = 4(\rho^{N_{\text{s}}} - \widetilde{\rho}^{N_{\text{s}}})^\dagger (\varphi^{N_{\text{s}}})^\ast.
\end{equation}
For $j = N_{\text{s}} - 1, N_{\text{s}} - 2, \ldots, 1$, we have
\begin{multline}
\label{eqn:genadj}
(\lambda^j)^\dagger = 4(\rho^j - \widetilde{\rho}^j)^\dagger (\varphi^j)^\ast 
+ (\lambda^{j+1})^\dagger K V_j^\text{prop}(\rho^j; \theta) K  \\
+ 2 \Re \Biggl[ \sum_{a} w^L_a \left( V_{j;aa}^\text{prop} (\rho^j;\theta) (-i \Delta t) \frac{\partial V^H_a}{\partial \rho^j} \right) w^R_a \Biggr] (\varphi^j)^\ast,
\end{multline}
where $w^L_a = \sum_{k} (\lambda^{j+1}_k)^\ast K_{ka}$ and $w^R_a = \sum_{b} K_{ab}$. There is only one remaining gradient to compute:
\begin{multline}
\label{eqn:thetagrad0}
\nabla_{\theta} L = \Re \sum_{j=0}^{N_s - 1} \sum_{k,a,b} (\lambda^{j+1}_k)^\ast K_{ka} V^\text{prop}_{j; aa} (\rho^j; \theta) (-i \Delta t) \\
 \cdot [\nabla_{\theta} V^C_{j,a}] K_{ab} \varphi^j_b
\end{multline}
With the representation $V^C_j = \theta_{j, :, :}$ mentioned above, we have $\partial V^C_{j,a} / \partial \theta_{\ell, c} = \delta_{j \ell} \delta_{a c}$, in which case
\begin{equation}
\label{eqn:thetagrad}
\partial_{\theta_{\ell,c}} L = \Re \sum_{k,b} (\lambda^{\ell+1}_k)^\ast K_{k c} V^\text{prop}_{\ell; cc} (\rho^{\ell}; \theta) (-i \Delta t) K_{cb} \varphi^{\ell}_b.
\end{equation}
The above induces an iterative algorithm, depicted in Figure \ref{fig:adjdiag}, to solve the minimization problem formulated above.  Suppose that at iteration $\ell$ we have a current best estimate $\theta^{(\ell)}$.  Then we carry out the following steps:
\begin{enumerate}
\item With the initial condition $\varphi^0$, we use (\ref{eqn:disceom}) to propagate forward in time and compute $\varphi$.  To evaluate $V^C$, we use our current iterate $\theta^{(\ell)}$.
\item Now that we have $\varphi$ (and by extension all electron densities $\rho^j$ for $j = 1, \ldots, N_s$), we initialize the adjoint system with the final condition (\ref{eqn:fincond}) and then use (\ref{eqn:genadj}) to propagate backwards in time and compute $\Lambda$.  Again, whenever we evaluate $V^C$ or its derivatives, we use $\theta = \theta^{(\ell)}$.
\item Having computed both $\varphi$ and $\Lambda$ in the above fashion, the gradient of our Lagrangian $L$ with respect to $(\varphi, \Lambda)$ is identically zero.  
\emph{Now we compute $\nabla_{\theta} L$ via (\ref{eqn:thetagrad}). Using it in a gradient-based optimization algorithm (such as Adam or L-BFGS), we compute the next iterate $\theta^{(\ell+1)}$.}  As we used (\ref{eqn:disceom}) and (\ref{eqn:genadj}) to compute $\varphi$ and $\lambda$, Lagrange multiplier theory guarantees that $\nabla_{\theta} L$ equals the total gradient of the loss with respect to $\theta$, i.e., differentiating through all propagation steps (\ref{eqn:disceom}).
\end{enumerate}
We iterate until the termination criterion $\| \nabla_{\theta} L (\theta^{\ell})\| < \epsilon$ is satisfied.  In this way, we find a (numerically approximate) critical point $(\varphi^\star, \Lambda^\star, \theta^\star)$ of the Lagrangian. 
As a final step, we verify numerically that the resulting trajectory of electron densities is sufficiently close to the reference data.  

\subsection{Direct Learning of a Memory-Dependent Functional: Adjoint Method}
\label{sect:directlearning}
The same framework just described enables us to directly learn $V^C$. The difference is that we explicitly allow $V^C$ to depend on a time history of electron densities:
\begin{equation}
\label{eqn:VC}
V^C = V^C(\rho^{k-M}, \ldots, \rho^{k-1}, \rho^k; \theta).
\end{equation}
The optimization problem is then nearly identical: minimize the loss (\ref{eqn:cost})
subject to (\ref{eqn:disceom}) and a prescribed initial segment $\{ \varphi^{-M}, \ldots, \varphi^0 \}$.  We reuse the Lagrangian (\ref{eqn:Lagrangian}) defined above and focus on taking variations with respect to $\varphi$ and $\varphi^\ast$.  We see that $\delta L_1$ is unchanged.  For $\delta L_2$, because $V^C$ depends on the history of $\rho$, our earlier calculations become more involved: for $a$ from $1$ to $N^2$,
\begin{multline*}
V^\text{prop}_{aa} = e^{-i \Delta t \bigl(V^{\text{ext}}_a + \frac{1}{2} V^H_a(\rho^j)  + V^C_a(\rho^{j-M}, \ldots, \rho^{j-1}, \rho^j; \theta) \bigr) }.
\end{multline*}
Varying $\varphi$ and $\varphi^\ast$, we find that
\begin{align*}
\delta &V^\text{prop}_{aa} = V^\text{prop}_{aa} (-i \Delta t) \biggl( \frac{1}{2} \frac{\partial V^H_a}{\partial \rho^j} \delta \rho^j \\
& \ + \sum_{m=0}^M \frac{\partial V^C_a}{\partial \rho^{j-m}} (\rho^{j-M},\ldots,\rho^j;\theta) \delta \rho^{j-m} \biggr) \\
 &= -i \Delta t \biggl( \frac{\partial V^H_a}{\partial \rho^j} ( (\varphi^j)^\ast \delta \varphi^j + (\delta \varphi^j)^\ast \varphi^j ) \\
 & \ + \sum_{m=0}^M \frac{\partial V^C_a}{\partial \rho^{j-m}} ( 2(\varphi^{j-m})^\ast \delta \varphi^{j-m} + 2 (\delta \varphi^{j-m})^{\ast} \varphi^{j-m} ) \biggr).
\end{align*}
In supplementary material, we show how the above calculation enables us to compute $\delta L_2$, set $\delta L = 0$ for all variations of $\varphi$ and $\varphi^\ast$, and derive the backward-in-time system that governs $\lambda$.  At step $j = N_s$, this system's final condition is
\[
(\lambda^{N_{\text{s}}})^\dagger = 4(\rho^{N_{\text{s}}} - \widetilde{\rho}^{N_{\text{s}}})^\dagger \varphi_{N_{\text{s}}}^\ast.
\]
For $j = N_{\text{s}} - 1, N_{\text{s}} - 2, \ldots, 1$, we evolve $\lambda$ backward in time via
\begin{multline}
\label{eqn:genadjlearn}
(\lambda^j)^\dagger = 4(\rho^j - \widetilde{\rho}^j)^\dagger \varphi_j^\ast 
+ (\lambda^{j+1})^\dagger K V(\theta) K  \\
+ \sum_{m=0}^M 4 \Re\biggl[(\lambda^{j+m+1})^\dagger 
 K  \frac{\partial V^\text{prop}(\rho^{j+m-M}, \ldots, \rho^{j+m};\theta)}{\partial \rho^{j}} \\
 \cdot K \varphi^{j+m} \biggr] 
  (\varphi^j)^\ast I_{j+m \leq N_{\text{s}}-1}.
\end{multline}
Here $I_A$ denotes the indicator function that equals $1$ if the condition $A$ is true and $0$ otherwise.  The gradient of $L$ with respect to $\theta$ is the same as in (\ref{eqn:thetagrad0}). However, because we are trying to learn a functional, we typically parameterize $V^C$ using a neural network.  In this case, $\theta$ represents the collection of all neural network weights.  To compute $\nabla_{\theta} V^C$, we employ standard automatic differentiation.   With these changes, the overall algorithm to minimize the loss (thereby training our $V^C$ model) remains the same as described above and as depicted in Figure \ref{fig:adjdiag}.

\section{Computational Details}
\label{sect:compdetails}
We consider two model two-electron systems.  The first is the hydrogen molecule $\htwo$ in a 50/50 superposition of two states, and the second is the scattering of an electron off a hydrogen atom.  For both system, all electrons are treated in two-dimensional real space, implying that the interacting wave functions $\Psi$ have an overall four-dimensional spatial dependence.  We begin by describing how we numerically solved the TDSE for each interacting system to generate ground truth 1-electron data.  We then describe additional steps taken to complete inversion and learning of $V^C$ for each system. Finally, we describe the density functionals from the literature that we use for a baseline comparison. 

\subsection{Solving the TDSE for \htwospace superposition}
\label{sect:htwo}
For \htwo, the domain of each coordinate is $[-5,5]$. We use $N = 128$ grid points with $\Delta x = 10/(N-1)$; we also use the soft-Coulomb  potential $V_\alpha(r)$ defined in Section \ref{sect:intro}.  We choose $\alpha = \Delta x$ so that bound states remain well-localized on our domain $\Omega$.  We use a 4th-order finite-difference stencil with Neumann (zero-flux) boundary conditions to discretize first partial derivatives; if $D_1$ is the resulting finite-difference matrix in one spatial dimension, we set $D_2 = -D_1^\dagger D_1$, effectively a 9-point stencil that retains both 4th-order accuracy and negative-semidefiniteness.  By using the highly sparse $D_2$ in a tensorial fashion, we can compute the action of the four-dimensional Laplacian on a trial wave function, here represented as a vector of length $N^4 = 128^4$.  Note that the potential is a diagonal matrix in real space, hence simple to incorporate.  Nuclei were separated by $1.4$ a.u.
We then leverage the matrix-free, large-scale Krylov-Schur eigenvalue solver in SLEPc\cite{SLEPc} to compute the first 14 eigenstates of $\htwo$ to a tolerance of $10^{-10}$.  With SLEPc operating on a GPU-enabled PETSc\cite{petsc-web-page,petsc-efficient} back end, the calculation fits on one Nvidia H200 GPU.

With this scheme, we find that the ground state energy of this model of $\htwo$ is $E_0 = -3.721$ Ha.  As compared with the usual reference value for $\htwo$ in three-dimensional space, this represents significantly more confinement of electron density near the nuclei. This is by design---we could have chosen the softening parameter $\alpha$ to reproduce the ground state energy $-1.174$ Ha, but we prioritized keeping electron density away from the boundary of the domain to reduce numerical artifacts that arise later in the inversion and learning process.

We then formed two different 50/50 superpositions of the ground state $\Psi_0$ with an excited state: $(\Psi_0 + \Psi_8)/\sqrt{2}$ and $(\Psi_0 + \Psi_{10})/\sqrt{2}$.  These superpositions were chosen because $\Psi_8$ and $\Psi_{10}$ have the same total spin as $\Psi_0$, and also because the corresponding energies are well-separated from $E_0$, i.e., $E_8 = -3.030$ Ha and $E_{10} = -2.911$ Ha.  Initializing in a superposition of the form $(\Psi_0 + \Psi_J)/\sqrt{2}$, the exact time-evolution is simply
\[
\Psi(t) = \frac{1}{\sqrt{2}} \left( \Psi_0 e^{-i E_0 t} + \Psi_J e^{-i E_J t} \right).
\]
We compute this on a temporal grid with $\Delta t = 0.005$ a.u. for $N_s = 4000$ time steps (excluding $t=0$).  From each  $\Psi(t)$, it is straightforward to compute the 1-electron density $\widetilde{\rho}(t)$, its time-derivative $\dot{\widetilde{\rho}}(t)$, and the 1-electron current $\widetilde{\mathbf{j}}(t)$.  We save these quantities to disk for the two superpositions mentioned above.  We use the first $N_t = 2000$ steps of each trajectory for training, and reserve the rest for our test sets.

\subsection{Solving the TDSE for H + e$^-$ scattering}
\label{sect:scattering}
For the scattering problem, the domain of each coordinate is $[-2 \pi, 2 \pi)$.  We use $N = 128$ grid points but now exclude the right-end point; we set $\Delta x = 4 \pi/N$ with two sets of grid points $x_j^{1}, y_j^{1} = -2 \pi + j \Delta x$  and $x_j^{2}, y_j^{2} = -2 \pi + (\Delta x)/2 + j \Delta x$, both for $j = 0, \ldots, 127$.  By using these staggered grids to treat the coordinates of each electron, we enable use of the Coulomb potential $V_0(r) = 1/r$ with no softening.    We take two further steps: first, we place the H atom at a location equidistant from the nearest point on either grid; the particular choice is $(x_H, y_H) = (2.75 \Delta x, 0.25 \Delta x) \approx (0.27, 0.0245)$.  This ensures that electrons on either grid feel an equal electron-nuclear potential.  Second, we initialize our system in the symmetric superposition
\begin{equation}
\label{eqn:psisuperpos}
\Psi = \frac{1}{\sqrt{2}} \left( \Psi_0(\br^1) \otimes \Psi_{WP}(\br^2) + \Psi_{WP}(\br^1) \otimes \Psi_{0}(\br^2) \right),
\end{equation}
where $\br^1 = (x^1, y^1)$, $\br^2 = (x^2, y^2)$, $\Psi_0(\br^j)$ is the numerically obtained ground state of the $H$ atom on grid $j$, and
\begin{multline*}
\Psi_{WP}(x,y) = \sqrt{\frac{5}{\pi}} \exp \bigl( -2.5((x-x_0)^2 + (y-y_0)^2) \\
+ i(p_x(x-x_0) + p_y(y-y_0)) \bigr)
\end{multline*}
is a Gaussian wave packet representing the incident electron centered at $(x_0,y_0)$ with initial momentum $(p_x,p_y)$.  Particular choices considered here are $(x_0,y_0) = (2.25, 0)$ with $(p_x,p_y) = (-3,0)$, $(-2.75,0)$, and $(-3.25,0)$.  
The combination of the above steps restores exchange symmetry to the extent allowed in our staggered grid formulation.  Note that (\ref{eqn:psisuperpos}) is the higher-dimensional analogue of the superposition studied in prior
work.\cite{PhysRevA.101.050501}

Once we have specified the interacting system's initial state, we step the four-dimensional TDSE forward in time via operator splitting. The wave function is represented as a vector of length $N^4 = 128^4$.  Thus the exact kinetic propagator (unlike the potential propagator) involves the matrix exponential of a non-diagonal matrix of size $N^4 \times N^4$.  We found through experimentation that sparse finite-difference approaches fail to maintain accuracy over long integration times and thus arrived at a pseudospectral approach.  

Since the Laplacian is diagonal in Fourier space, the kinetic propagator reduces to multiplication by a diagonal matrix, sandwiched by one FFT (fast Fourier transform) and one inverse FFT.  On benchmark problems with exact solutions, this approach yields more than $8$ digits of accuracy. As operator splitting involves only unitary matrices, normalization of $\Psi$ is preserved to machine precision throughout propagation.  The only downside of this approach is the implied periodic boundary conditions.  However, we have chosen system parameters, including a final time $T$, such that the wave function always remains negligible near the boundary of the domain.

We propagate forward in time for $N_s = 8000$ steps using a time step of $4.8 \times 10^{-21}$ seconds or $1.984 \times 10^{-4}$ a.u.  Every $10$ time steps, we save to disk the 1-electron density $\widetilde{\rho}(t)$, its time-derivative $\dot{\widetilde{\rho}}(t)$, and the 1-electron current current $\widetilde{\mathbf{j}}(t)$.

\subsection{Inversion and learning for \htwo}
\label{sect:ilh2}
For a 2-electron system described by a single KS orbital $\varphi$, one can employ the Madelung transformation $\varphi = \sqrt{ \rho / 2} e^{i \zeta}$ to derive the continuity equation
\begin{equation}
\label{eqn:qhd}
\odiv( \rho \nabla \zeta ) = -\dot{\rho}.
\end{equation}
If $\rho$ and $\dot{\rho}$ are known, one can solve for $\zeta(\bx, t)$.  This enables determination of $v^S$ and then $v^C$ at each point in spacetime.\cite{ullrich2011time}   Note that if the 1-electron current $\mathbf{j}$ is known, one could employ $\dot{\rho} + \odiv(\mathbf{j}) = 0$ to write
\begin{equation}
\label{eqn:qhdj}
\odiv( \rho \nabla \zeta ) = \odiv(\mathbf{j}).
\end{equation}
Thus we have inversion pathways from either $(\rho, \dot{\rho})$ or $(\rho, \mathbf{j})$ to the correlation potential $v^C$.   Because Madelung was an originator of quantum hydrodynamics (QHD), and because it is useful to have a label for this inversion procedure, we refer to it as the QHD approach.  For molecular systems, $\rho$ will decay to zero far from the nuclei, rendering both (\ref{eqn:qhd}) and (\ref{eqn:qhdj}) ill-posed.   In prior work, this problem has been mentioned but not completely resolved.\cite{Ruggenthaler2015,DLFM21}

For our $\htwo$ system, we begin the inversion and learning process via QHD inversion from $(\widetilde{\rho}, \dot{\widetilde{\rho}})$ to $V^C$ on our spacetime grid.  Using this $v^C$ as an initial guess, we then employ PDE-constrained optimization with our adjoint method (see Section \ref{sect:invadj}) to generate a refined $\widetilde{V}^C$.  To compute the kinetic propagator $K$, we compute the exact matrix exponential using the finite-difference matrix $D_2$ defined above together with the Kronecker-sum exponential identity. Here we add to our loss a smoothness penalty $\lambda_S \| \nabla V^C \|^2$, with $\nabla$ discretized using $D_1$ described above.  We find that $\lambda_S = 10^{-8}$ is sufficient for accurate results.   The adjoint method computes gradients of the loss; we must also choose an optimization method that uses those gradients to step towards a minimizer.  Here we employ the Adam optimizer with initial learning rate of $10^{-2}$, exponentially decayed to decrease by one order of magnitude every $2000$ steps.

Having solved for $\widetilde{V}^C$, we reshape it into a matrix $V$ of size $4000 \times 128^2$ and then compute its singular value decomposition (SVD) $V = U \Sigma B^\dagger$.  Based on the decay of singular values, we select the first $256$ rows of $B^\dagger$ (which in the thin SVD has size $4000 \times 128^2$) and save these as $B_0^\dagger$.  This $B_0$ comprises a reduced-dimensional, data-driven basis in which we can represent $V^C$.

Carrying out the same procedure for snapshots of the electron density $\rho$, we obtain a matrix $R_0^\dagger$ of size $4 \times 128^2$.   Because $\rho$ is smoother in space and time than $V^C$, we can compress its dimensionality much more.  All steps are carried out on each superposition trajectory in turn.  

The next step in our workflow is to train a $V^C$ model with explicit memory-dependence.  As we have access both to ground truth $\widetilde{\rho}$ and $\widetilde{V}^C$, this learning can proceed offline: propagation through the TDKS equations is not necessary, and thus the adjoint-based optimization loop can be bypassed.  We use a neural network model that takes as input a reduced-dimensional representation of $M$ total snapshots of the electron density.  That is, given
\begin{equation}
\label{eqn:rhohist}
\rho^j_M = (\rho^{-M+1+j}, \ldots, \rho^j),
\end{equation}
reshaped to be of size $M \times N^2$, we can multiply on the right by $R_0$ to compress its dimensionality to $M \times 4$, which we reshape into a vector of length $4M$.  Given this vector, our neural network architecture is simple: we have four, dense, feedforward layers each consisting of the transformation $\bz \mapsto \phi( \bz W^{\ell} + \bbee^{\ell})$ where $\phi(z) = z I_{z \geq 0}$ is the rectified linear unit.  Our dense layers have $256$ units each.  The output from the final layer is passed through a final linear transformation $\bz \mapsto \bz W^5 + \bbee^5$, right-multiplied by $B_0^\dagger$, and finally reshaped to size $128 \times 128$ to yield $V^C[\rho^j_M; \theta]$.

Let us define the single-trajectory density-weighted loss by
\begin{equation}
\label{eqn:denweightloss}
L_{\widetilde{\rho}}(\theta) = \sum_{j=M-1}^{N_t} \sum_{a b} \widetilde{\rho}^j_{ab} (V^C[\widetilde{\rho}^j_M; \theta]_{a b} - \widetilde{V}^C_{a b} )^2 .
\end{equation}
But for the prefactor of $\widetilde{\rho}^j_{ab}$, this would be the sum of squared errors between the $V^C$ predicted by our neural network and the $\widetilde{V}^C$ we obtained from PDE-constrained inversion.  The prefactor encourages the neural network to focus on matching $\widetilde{V}^C$ in regions of the domain where there is more electron density.  Here $N_t$ is the number of steps used for training.  Note also that the sum begins at $j=M-1$,  the smallest value of $j$ such that (\ref{eqn:rhohist}) is well-defined.

We sum this single-trajectory loss over both superposition trajectories and minimize it using $20000$ steps of the Adam optimizer followed by $120000$ steps of L-BFGS.  All gradients are computed via automatic differentiation. 

Once we have an \emph{offline trained} $V^C$ model, we refine it by incorporating it into our adjoint-based optimization procedure.  At each time step, we form the history (\ref{eqn:rhohist}) using the propagated TDKS states rather than the ground truth $\widetilde{\rho}$ snapshots.  We apply the dimensionality reduction technique described above to compress this time history into a vector of length $4M$, pass this into the neural network, and use the resulting $V^C$ to propagate to the next step.  We use the direct learning adjoint method from Section \ref{sect:directlearning} to compute gradients of the loss (\ref{eqn:cost})---averaged over both training trajectories---with respect to the neural network parameters $\theta$.  Here we employ $10000$ steps of the Adam optimizer followed by $10000$ steps of L-BFGS.  We call the resulting $V^C$ model our \emph{fine-tuned} neural network model.

We must mention one final subtlety regarding phases.  When we propagate the TDKS equations using a memory-based $V^C$ model, we must begin propagation at time step $j = M-1$.  Thus $\varphi^{M-1}$ is needed to begin propagation.  We have found empirically that if we fix $\varphi^{M-1}$ equal to the QHD result (the first step in the procedure above) but change $V^C$ (either during offline or adjoint-based training), we obtain inaccurate results.  This is because $V^C$ at time $M-1$ and the phase $\zeta^{M-1}$ are closely interrelated, as the equations of QHD inversion make clear.  In order to overcome this obstacle, when we train our models for $\htwo$, we optimize over both $\theta$ and the phase $\zeta^{M-1}$.  To ensure that this training does not deviate substantially from the physics, we add to either (\ref{eqn:denweightloss}) or (\ref{eqn:cost}) a penalty of the form
\begin{equation}
\label{eqn:poissonpen}
\sigma(\zeta^{M-1}) = \lambda_P \| \odiv( \widetilde{\rho}^{M-1} \nabla \zeta^{M-1} ) + \dot{\widetilde{\rho}}^{M-1} \|^2,
\end{equation}
with $\odiv$ and $\nabla$ discretized using the $D_1$ matrix described above. We choose $\lambda_P = 10^3$. The gradient of (\ref{eqn:poissonpen}) with respect to $\zeta^{M-1}$ is simple.  Another contribution to this gradient arises because $\zeta^{M-1}$ influences $\varphi^{M-1}$, the initial condition for our TDKS propagation; we compute this contribution via the adjoint method.  

\subsection{Direct learning for H + e$^{-}$ scattering}
\label{sect:directscattering}
To train a memory-based $V^C$ model via direct learning, we use a different kind of neural network than above.  Because we avoid solving for spacetime values of $V^C$ in this approach, we have no ground truth $\widetilde{V}^C$ to use for dimensionality reduction.  Thus our neural network architecture must be chosen more carefully to avoid blow-up in the number of parameters.

For this reason, we have explored convolutional neural network (CNN) architectures.  Our model's first layer takes as input a sequence of $M$ snapshots of the density, in the form of an $N \times N \times M$ tensor.  We treat the last axis as the channel dimension.  From there, each layer of the network applies sets of three-dimensional convolutional kernels in an effort to \emph{encode} information from the input into low-dimensional objects.  Our CNN uses four convolutional layers to gradually transform the input from $128 \times 128 \times M$ to shape $8 \times 8 \times 128$.  After applying a dense layer vertically in the channel/feature dimension, to mix these latent features, we apply four additional convolutional layers to transform or \emph{decode} back to an output of size $128 \times 128 \times 1$.  Dropping the third axis yields our $v^C$ output.

Each layer of the network includes a linear convolution step followed by a nonlinear activation function, here chosen to be the scaled exponential linear unit.\cite{Klambauer2017}  The total parameter count is 361,697, far less than that of a dense feedforward network with the same input and output dimensions.  The overall architecture we have used is known as a U-net\cite{ronneberger2015u}, minus skip connections, which we found do not improve predictive power in this case.  Essentially, our network learns the encoding and decoding that was hardcoded (via the SVD matrices $R_0$ and $B_0$) in the model described above.

For the scattering problem, we employed QHD inversion to compute initial phases $\zeta^{M-1}$.  Subsequently, we did not reoptimize these phases.  For 2D TDKS propagation, we used an FFT-based approach to compute the kinetic propagator, matching the approach used to solve the 4D TDSE for the scattering problem.  To compute the Hartree potential, however, we used free space or natural boundary conditions, thus avoiding periodic artifacts.

To train, we used the trajectories with momenta $(p_x,p_y) = (-2.75,0), (-3.25,0)$, reserving for the test set the $(-3,0)$ trajectory.  Training proceeded via 100000 steps of L-BFGS optimization.  Because the training data was saved at a time step of $\Delta t = 1.984 \times 10^{-3}$ a.u. (every 10th step of the corresponding TDSE simulation), we use that as our time step for TDKS forward and adjoint propagation.  While the entire trajectory has length $N_s = 800$ steps, we use only $N_t = 400$ steps for training.

\begin{figure}[t]
\includegraphics[width=3.37in]{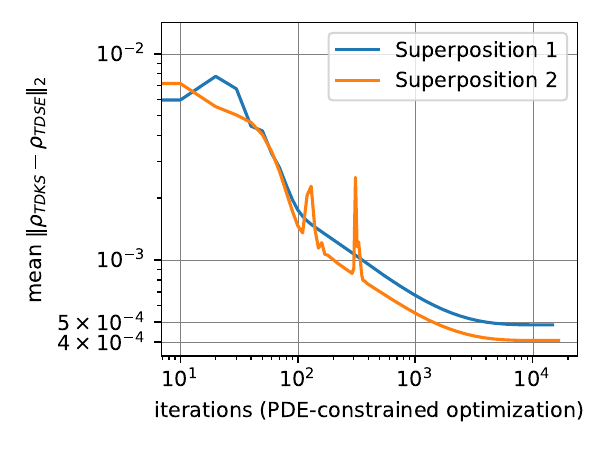}
\caption{Applying the adjoint-based inversion procedure from Section \ref{sect:invadj} to the training data described in Section \ref{sect:htwo}, we obtain the above mean $L_2$ errors as a function of iteration count.  The final $\widetilde{V}^C$'s found through this procedure, when used to solve the TDKS equations for our $\htwo$ model, yield densities that match reference trajectories with time-averaged $L_2$ errors of roughly $5 \times 10^{-4}$ and $4 \times 10^{-4}$, respectively.}
\label{fig:inversionadjerr}
\end{figure}

\subsection{Density functionals for baseline comparison}
\label{sect:density_functionals}

Here we use standard abbreviations for LDA (local density approximation), ALDA (adiabatic LDA), GGA (generalized gradient approximation), PBE\cite{PBE}, and PRM.\cite{PRM} Note that the latter two acronyms are named after the authors of the functionals, with the PRM LDA correlation functional designed for 2D electron systems.\cite{PRM} These functionals were developed based on properties of ground state electron densities and are employed within the standard adiabatic approximation, taking no account of the past history of the density or initial state of the system. LDA functionals take as input only the value of the density on a spatial grid point; thus, they are considered completely local. The GGA PBE $v^X$ also takes as input the gradient of the density at a spatial grid point; thus, it is considered a semi-local density functional. 

We denote the three density functional models that we compare our results to as:
\begin{enumerate}
\item ALDA1: exact exchange $v^X$ plus LDA $v^C$,
\item ALDA2: LDA $v^X$ plus LDA $v^C$, and
\item GGA: PBE $v^X$ plus PRM (LDA) $v^C$.
\end{enumerate}
Note that our GGA density functional model includes the PBE GGA exchange potential $v^X$, but we have chosen the LDA PRM correlation potential for $v^C$, thinking that it may provide improved performance for these spatially 2D systems. In all settings, we use standard pylibxc/libxc implementations of these models in two spatial dimensions.\cite{lehtola2018recent}

\section{Results}
\label{sect:results}
\subsection{\htwo: invert-then-learn}
\label{sect:h2itl}
As described in Section \ref{sect:ilh2}, we begin with PDE-constrained optimization to generate ground truth values of the correlation potential $\widetilde{V}^C$ on our spacetime grid.  In Figure \ref{fig:inversionadjerr}, we show the results of using the Adam optimizer in conjunction with our adjoint method to compute gradients of the loss.  The metric here is
\begin{equation}
\label{eqn:meanL2error}
\frac{1}{N_t} \sum_{j=1}^{N_t} \left[ \sum_{a,b} ( \rho_{a,b}^j - \widetilde{\rho}_{a,b}^j )^2 (\Delta x)^2 \right]^{1/2}.
\end{equation}
The quantity inside the summation over time (index $j$) is the $L_2$ norm in continuous space  between the functions $\rho$ and $\widetilde{\rho}$.  We average this over time to compute each error plotted in Figure \ref{fig:inversionadjerr}.

Recall that we initialized our PDE-constrained optimization with the results of QHD inversion.  Thus the left endpoints on the two plotted curves tell us the mean $L_2$ propagation error (\ref{eqn:meanL2error}) when we use the $V^C$ from QHD inversion.  Even for a well-behaved two-electron system, QHD does not yield $V^C$ values that are accurate enough to train a machine learning model.  By applying PDE-constrained optimization, we reduce the mean $L_2$ error by roughly two orders of magnitude.

\begin{figure}[t]
\includegraphics[width=3in,clip,trim=10 10 10 10]{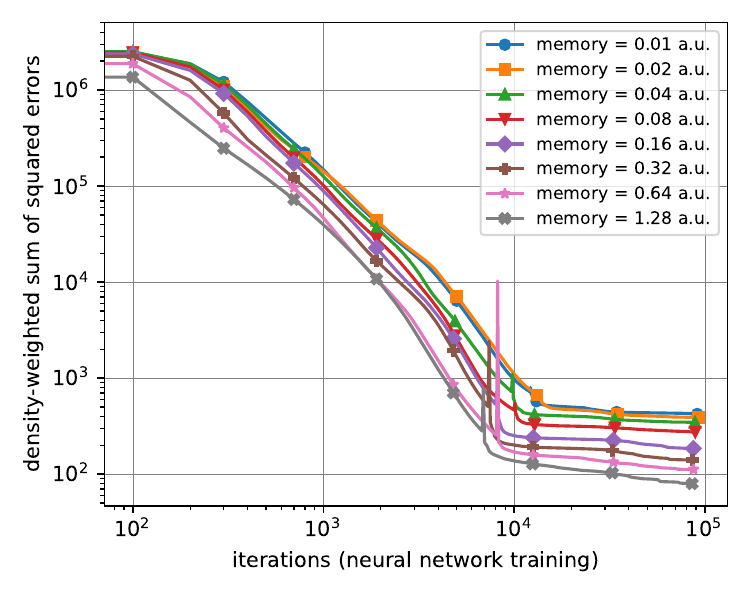}
\caption{Using the $\widetilde{V}^C$ values obtained from PDE-constrained optimization, we employ offline training---see Section \ref{sect:ilh2}---to learn a $V^C$ model for H$_2$ that depends on a time history of $M$ snapshots of the density $\rho$.  Here we study how the loss (\ref{eqn:denweightloss}) behaves during training as a function of $M\Delta t$, the memory length in a.u.  We  see that the final optimized loss monotonically decreases as a function of $M$.}
\label{fig:lossvsmemory}
\end{figure}

Having found $\widetilde{V}^C$, we use it for offline learning of memory-based $V^C[\rho^j_M; \theta]$ neural network models.  In Figure \ref{fig:lossvsmemory}, we show how the density-weighted sum of squared errors (\ref{eqn:denweightloss}) depends on memory $M \Delta t$ measured in atomic units.  Here the loss is summed over both superposition trajectories. By virtue of employing offline learning, in which we do not have to propagate through the TDKS equations on each optimization step, we can quickly train multiple long memory models.  Each model is trained on both superposition trajectories at once.  We find that increasing memory $M$ strictly decreases the final optimized value of the loss function.   For subsequent work in this paper, we choose the $M=256$ model, corresponding to a memory of $1.28$ a.u. In prior work on a memory-based model to propagate 1-electron reduced density matrices, we found that for $\htwo$ in two small basis sets, memory of $0.8$ to $13.0$ a.u. was required for accurate propagation; the value chosen here is in this window.\cite{BhatJMP}

\begin{figure}[t]
\includegraphics[width=3.37in]
{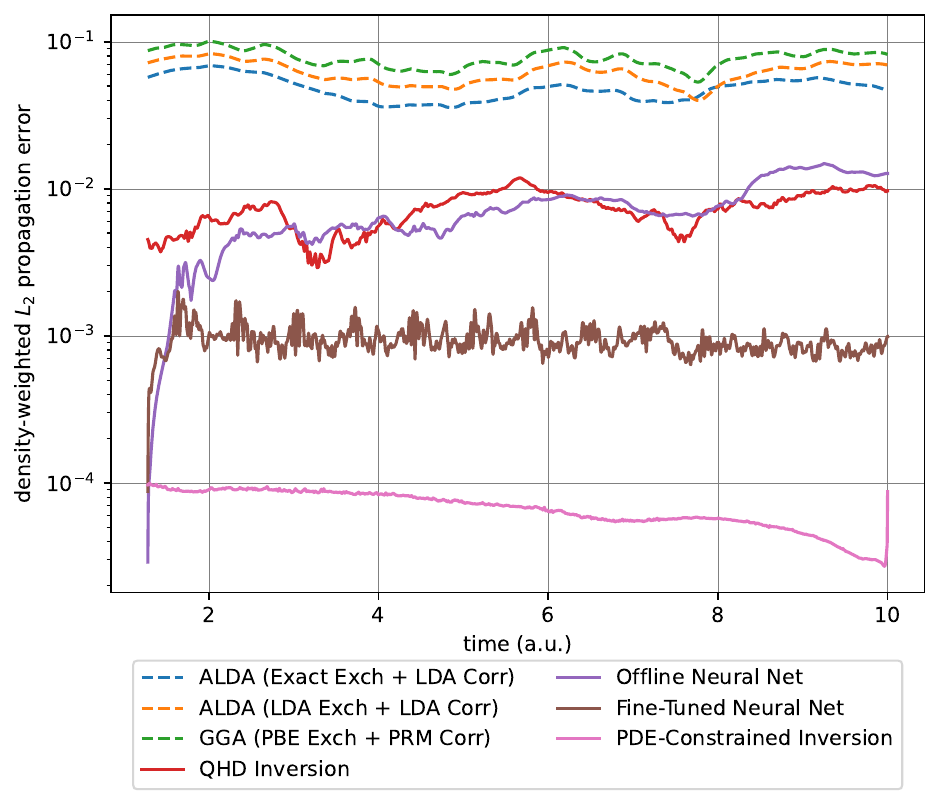}
\caption{Here we compare the density-weighted $L_2$ propagation error as a function of time for seven models for H$_2$: two inversion models we computed (QHD and PDE-Constrained Inversion), two neural network models (offline-trained and fine-tuned), and three baseline density functional models (ALDA1, ALDA2, and GGA).  This comparison is on the training set that spans $N_t = 2000$ points in time.  The best models, the fine-tuned neural network and PDE-constrained optimization, are respectively two and three orders of magnitude more accurate than the baselines.}
\label{fig:ftwb}
\end{figure}

As described in Section \ref{sect:ilh2}, we now fine-tune the $V^C[\rho^j_M; \theta]$ model by adapting the direct learning method from Section \ref{sect:directlearning}.  We then run a training set comparison including all models we have trained along with three standard density functionals from the literature.  For our model of $\htwo$, we propagate the TDKS system with each of the models.  For all models, we then compute the density-weighted $L_2$ propagation error:
\begin{equation}
\label{eqn:denmeanL2error}
E_j = \left[ \sum_{a,b} \frac{1}{2} \widetilde{\rho}_{a,b}^j ( \rho_{a,b}^j - \widetilde{\rho}_{a,b}^j )^2 (\Delta x)^2 \right]^{1/2}.
\end{equation}
At each point in time, the integral of $\widetilde{\rho} / 2$ over space equals unity; hence this quantity is a valid probability density.  Thus we can interpret (\ref{eqn:denmeanL2error}) as the expected value of the $L_2$ propagation error, where the expectation is taken with respect to the probability density in question.  

In Figure \ref{fig:ftwb}, we plot $E_j$---averaged over the two superposition trajectories---for four $V^C$ models developed in this paper along with the three baseline density functional comparison models.  All models have stable errors over the training window.  Starting from the density functional baseline models (all of which perform similarly), we pick up one order of magnitude of accuracy with either the QHD $V^C$ or the offline-trained neural network $V^C$, another order of magnitude with the fine-tuned neural network, and a further order of magnitude with the  $\widetilde{V}^C$ from PDE-constrained optimization.

\begin{figure}[t]
\includegraphics[width=3.37in]{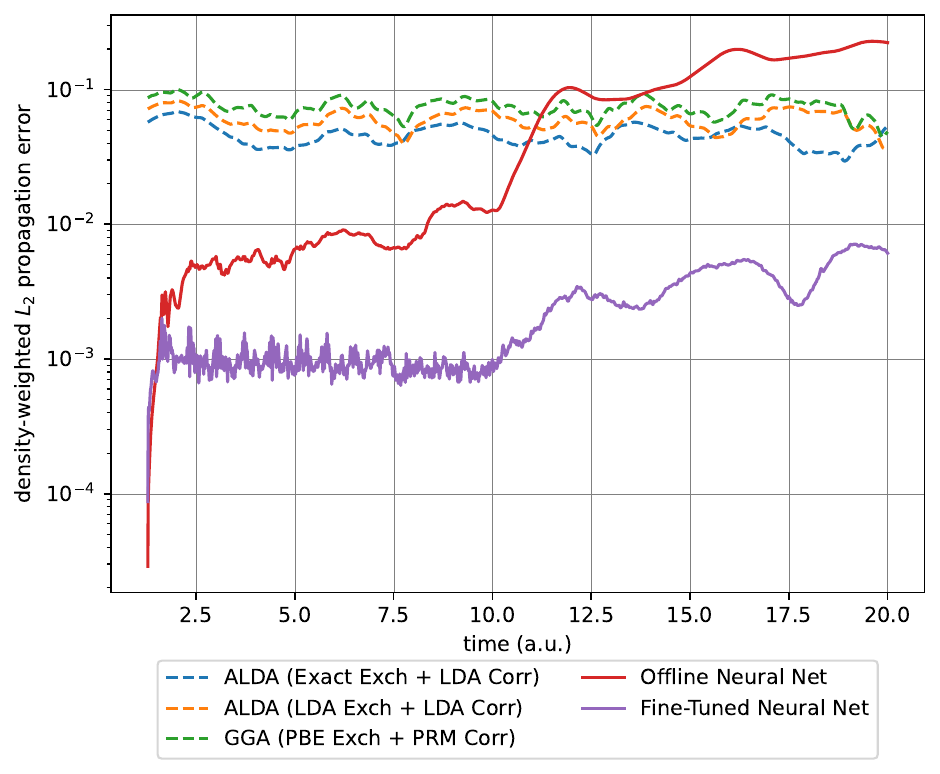}
\caption{Comparison of the density-weighted $L_2$ propagation error as a function of time for five models applied to H$_2$: two neural network models (offline-trained and fine-tuned) and three baseline density functionals (ALDA1, ALDA2, and GGA).  This comparison is on the entire data set; results after $t = 10$ a.u. are test set results.  The fine-tuned neural network retains an advantage over baseline models even when extrapolating $2000$ points beyond its training window.}
\label{fig:extrap}
\end{figure}

Three points emerge from Figure \ref{fig:ftwb}.  First, the offline-trained neural network does not capture enough signal from its training data, the $\widetilde{V}^C$ from PDE-constrained optimization, to match its performance. Second, though the fine-tuned neural network improves on the offline-trained model, there is still significant room for improvement.  Third, the baseline density functional models lead to objectively inaccurate results for this 50/50 superposition problem.

As described above, we trained on the first $N_t = 2000$ time steps of our trajectories, reserving the final $2000$ time steps as a test set.  In Figure \ref{fig:extrap}, we show test set results for our two neural network models (offline-trained and fine-tuned).  We once again use the density-weighted metric (\ref{eqn:denmeanL2error}), this time plotted from $j = 1$ to $j = N_s = 4000$.  The time $10$ a.u. is the point at which results switch from training to test; as shown, the errors of our neural network models increase after this point.  The offline-trained model performs poorly relative to the  baselines.  However, the fine-tuned neural network maintains reasonable accuracy up until the final time of $20$ a.u.  This demonstrates that our fine-tuned neural network has learned at least some features of the true $V^C$ potential.

\begin{figure}[h!]
\includegraphics[width=3.1in]{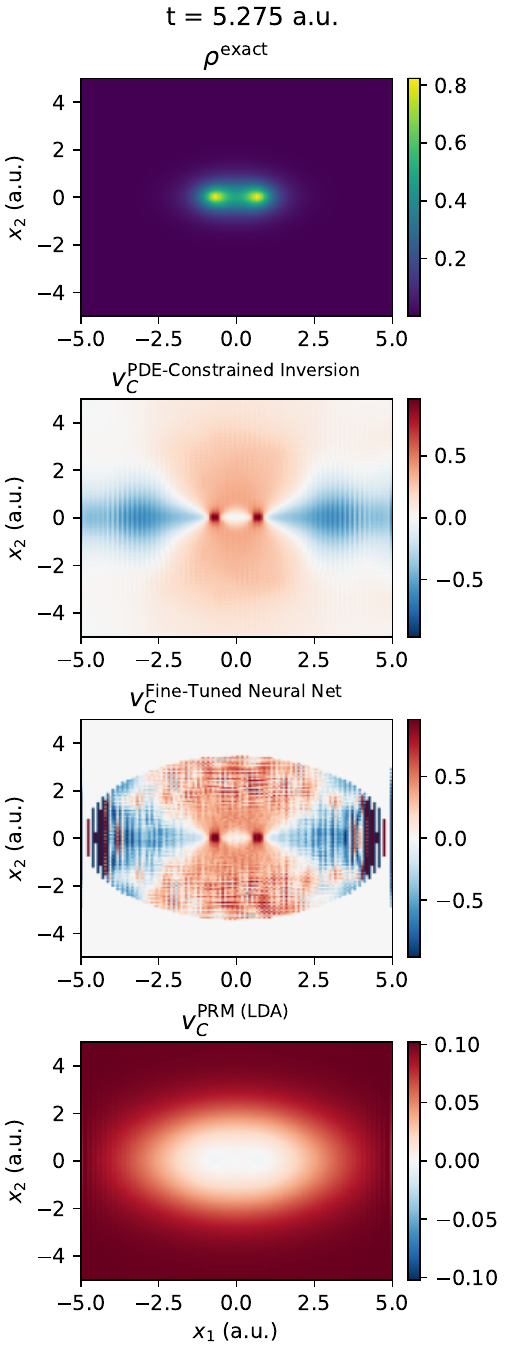}
\caption{At $t = 5.275$ a.u., we plot the reference density for H$_2$ along with three correlation potentials.  The neural network captures dynamical step features of the PDE-constrained $\widetilde{V}^C$ that are missed entirely by the nonnegative PRM $V^C$. For further details, consult Section \ref{sect:h2itl}.}
\label{fig:htwocomparison}
\end{figure}

In Figure \ref{fig:htwocomparison}, we visualize the potentials at time $t = 5.275$ a.u., a point in the middle of the training window.  We plot, from top to bottom, the reference density $\widetilde{\rho}$, the $\widetilde{V}^C$ correlation potential obtained from PDE-constrained inversion, the fine-tuned neural network potential $V^C$ (memory-based), and the adiabatic PRM (LDA) $V^C$.  The PDE-constrained $\widetilde{V}^C$ displays sharp gradients near the nuclei where the potential swings quickly from near $-0.5$ to near $0.5$, analogous to the step and peak features discussed in prior work.\cite{DLFM21}  In contrast, the PRM $V^C$ is overly smooth and entirely nonnegative.  Note that when we plot the fine-tuned neural network potential $V^C$ here, we mask out the region where $\widetilde{\rho} \leq 10^{-4}$.  In this low-density region, the neural network $V^C$ displays oscillations that play no role in propagation.

\begin{table}[t]
\centering
\begin{tabular}{lccc}
\hline
\textbf{Method} & \textbf{Loss} & \makecell{\textbf{Mean}\\\textbf{$L_2$ Error}} & \makecell{\textbf{Mean}\\\textbf{Density-Weighted}\\\textbf{$L_2$ Error}}\\
\hline
ALDA1 & $2.25\times 10^{3}$ & $2.98\times 10^{-1}$ & $1.69\times 10^{-1}$ \\
ALDA2 & $2.29\times 10^{3}$ & $3.00\times 10^{-1}$ & $1.73\times 10^{-1}$ \\
GGA & $2.42\times 10^{3}$ & $3.08\times 10^{-1}$ & $1.83\times 10^{-1}$ \\
U-Net & $5.27\times 10^{-1}$ & $4.81\times 10^{-3}$ & $1.76\times 10^{-3}$ \\
\hline
\end{tabular}
\caption{For the H + e$^-$ scattering problem, we compare loss and error metrics across different methods on the test set trajectory.  We include in this table three baseline density functional methods explained fully in Section \ref{sect:h2itl}; in summary, ALDA1 involves exact exchange and LDA correlation, while ALDA2 uses LDA exchange and LDA correlation. All baseline methods are roughly two orders of magnitude worse than the trained U-Net $V^C$.}
\label{tab:scatteringtest}
\end{table}

\begin{figure}[t]
\includegraphics[width=3.37in]{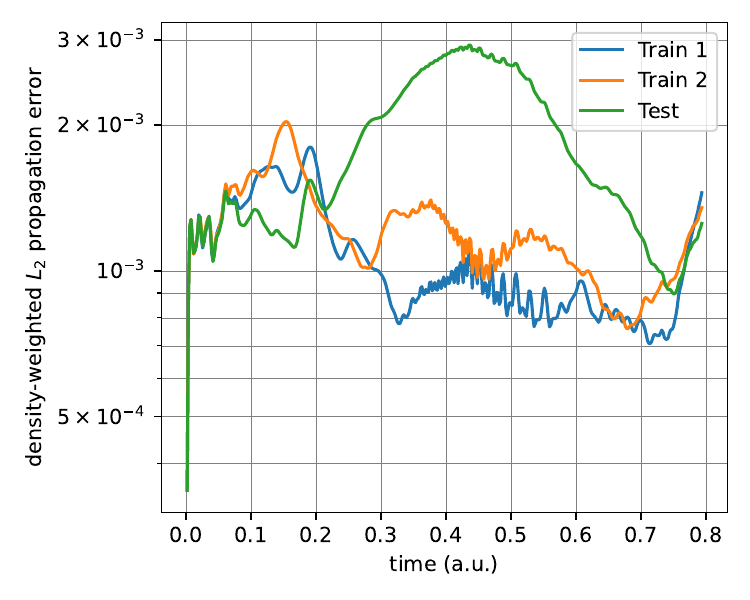}
\caption{For the H + e$^-$ scattering problem, we use the adjoint-based method from Section \ref{sect:directlearning} to train one U-Net $V^C$ model on two trajectories. Using this trained U-Net, we propagate with three different initial conditions corresponding to the training and test set trajectories.  Plotted here is the density-weighted $L_2$ propagation error between (i) densities obtained from TDKS propagation with the U-Net $V^C$ and (ii) reference densities computed from accurate TDSE solutions.  The test set error remains less than $3 \times 10^{-3}$ throughout propagation, indicating close agreement between densities.}
\label{fig:scatteringerror}
\end{figure}

\subsection{H + e$^{-}$ scattering: direct learning}
We train the U-Net model with memory $M = 2$ starting from a random initialization for all neural network weights.  The initial value of the loss (\ref{eqn:cost}), averaged across two training trajectories, is $2.23 \times 10^{3}$.  After training, the loss is $4.6 \times 10^{-1}$.  Since the loss is half the sum of squared errors, this already indicates excellent training fit.  Thus for the present work we did not explore U-Net models with memory $M > 2$.

In Figure \ref{fig:scatteringerror}, we plot the density-weighted $L_2$ propagation error (\ref{eqn:denmeanL2error}) over time for both training trajectories as well as the test set trajectory.  Note that errors remain below $3 \times 10^{-3}$ during the entire propagation window.  The test set error peaks when the incident electron interacts most directly with the hydrogen atom.

\begin{figure*}[t]
\includegraphics[width=6.69in]{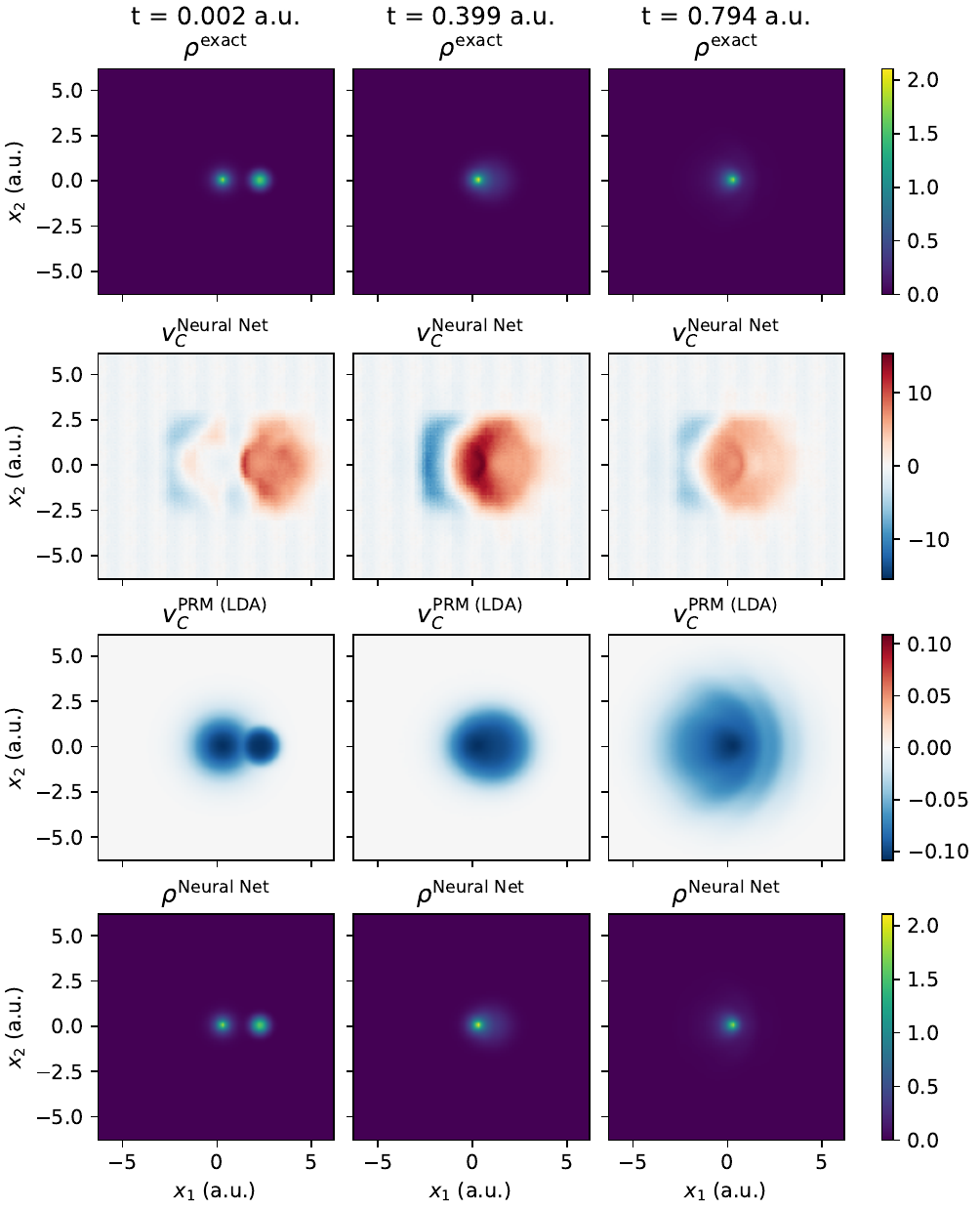}
\caption{At three key times during propagation (beginning, middle, and end), reference densities (top row, computed via solution of the TDSE) closely agree with densities produced by propagating TDKS with our trained U-Net $V^C$ (bottom row).  The middle two rows compare the U-Net $V^C$ against those computed via the PRM (LDA) functional.  The U-Net captures dynamical step/peak features; the LDA is incorrect both in magnitude and in sign-definiteness (no positive values).}
\label{fig:bigkahuna}
\end{figure*}

In Figure \ref{fig:bigkahuna}, we display the time-evolution of densities and correlation potentials at three key points in time. Note that these results are for the \emph{test set} trajectory.  From left to right, each column shows results at $t = \Delta t = 1.984 \times 10^{-3}$ a.u. (just after propagation begins), $t = 201 \Delta t = 0.399$ a.u. (halfway through propagation), and $t = 400 \Delta t = 0.794$ a.u. (when propagation ends).  The top and bottom rows show excellent agreement between reference densities and those produced by propagating with the U-Net $V^C$. 

The middle two rows capture striking differences between U-Net and PRM (LDA) correlation potentials.   The single doubly occupied KS orbital for this system must somehow be able to capture the dynamics of both the incident electron and the electron bound to the H atom.  In order for this KS orbital to evolve in time with sufficient accuracy so that the density matches that of the interacting system, the true $V^C$ must have large-magnitude steps/peaks.
Though the U-Net was never trained using ground truth values of the correlation potential---not even those that we might obtain via QHD---we see evidence of dynamical steps and peaks, where the potential changes sign and swings from high to low magnitudes.  In contrast, the PRM (LDA) $V^C$ is two orders of magnitude smaller and is also completely nonpositive (i.e., either negative or zero).  

Note in particular the especially large magnitude step in the U-Net $V^C$ at time $t = 0.399$ a.u., corresponding to the time at which the test set error peaks in Figure \ref{fig:scatteringerror}.  Though this error may seem relatively large on the scale of Figure \ref{fig:scatteringerror}, when we examine the densities at $t = 0.399$ a.u. in Figure \ref{fig:bigkahuna}, they are indistinguishable.

In Table \ref{tab:scatteringtest}, we quantify test set errors for the trained U-Net and the three baseline models mentioned in Section \ref{sect:h2itl}.  In either propagation error metric, the U-Net correlation potential yields densities that are roughly $100$ times more accurate than baseline methods.
Note also that the mean density-weighted $L_2$ error for the U-Net is less than half the mean $L_2$ error, indicating that its errors are less pronounced when electron density is higher.

\section{Conclusion} 
There are three main conclusions we draw from this work and the results.  First, one of the most important ingredients to accurate inversion and learning is the quality of the training data.  Without accurate reference densities, time-derivatives of these densities, and/or currents, inversion and learning are simply not possible.  Here, by paying close attention to the numerical solution of interacting systems featuring two electrons both in two spatial dimensions, we generated highly accurate training data to begin the inversion and learning process.

In a similar vein, we can see that QHD is too inaccurate to generate reference values of the correlation potential $V^C$ that are suitable for learning correlation functionals.  One can, however, use QHD as a starting point to solve for more accurate references values $\widetilde{V}^C$ via PDE-constrained optimization.  The first adjoint method we described solves this problem and, for $\htwo$, yields the best propagation error among all methods considered.

The second conclusion is that, despite using only two training trajectories, we have learned correlation functionals that feature low test set error either when extrapolating to unseen times (for \htwo) or extrapolating to initial conditions not seen in the training data (for hydrogen-electron scattering).  We also see that our trained functionals yield substantially lower TDKS propagation error (measured in either raw or density-weighted $L_2$ norms) than standard functionals in two dimensions.  This is true even in the direct learning case, where the neural network has never seen ground truth values of the correlation potential. We conjecture that if we were to retrain our models on much larger training sets,  we would be able to learn a memory-dependent correlation functional that yields accurate propagation well outside the training set.

Finally, our work points the way towards several areas for improvement.  This includes (i) improving the neural network architecture in the invert-then-learn approach, so that we reduce the propagation error gap between this approach and PDE-constrained optimization, (ii) studying methods to initialize propagation when using correlation functionals with long memory (e.g., how do we obtain the initial segment of electron densities to begin propagation?), and (iii) understanding how to incorporate known symmetries and exact conditions into our model functionals.

\section{Supplementary Material}
In supplementary material, we give mathematical derivations of the adjoint systems (\ref{eqn:genadj}) and (\ref{eqn:genadjlearn}).  This includes full details regarding variations of the Lagrangian (\ref{eqn:Lagrangian}) in both the inversion and direct learning cases.

\section{Acknowledgments} 
We are grateful for discussions of this work with N. Maitra, V. Gavini, B. Kanungo, and P. Zimmerman. This work was supported by the U.S. Department of Energy, Office of Science, Office of Advanced Scientific Computing Research and Office of Basic Energy Sciences, Scientific Discovery through Advanced Computing (SciDAC) program under Award Number DE-SC0026088 (TDKS inversion, machine learning of correlation potentials), and by the Office of Naval Research, Grant Number W911NF-23-1-0153 (4D TDSE implementation and optimization). This research used resources of the National Energy Research Scientific Computing Center (NERSC), a U.S. Department of Energy Office of Science User Facility located at Lawrence Berkeley National Laboratory, operated under Contract No. DE-AC02-05CH11231 using NERSC award BES-m5214. We also acknowledge computational time on the Pinnacles cluster, including CENVAL-ARC GPU nodes, at UC Merced, supported by NSF awards OAC-2019144 and OAC-2346744.

\section{References}
\bibliography{isborn,ciss,bhat,maitra}

\end{document}